# Stability and electronic properties of small boron nitride nanotubes

Zhuhua Zhang, Wanlin Guo[a], and Yitao Dai

*Institute of Nano Science, Nanjing University of Aeronautics and Astronautics,*

*Nanjing 210016, People's Republic of China*

We report the stability and electronic structures of the boron nitride nanotubes (BNNTs) with diameters below 4 Å by semi-empirical quantum mechanical molecular dynamics simulations and *ab initio* calculations. Among them (3,0), (3,1), (2,2), (4,0), (4,1) and (3,2) BNNTs can be stable well over room temperature. These small BNNTs become globally stable when encapsulated in a larger BNNT. It is found that the energy gaps and work functions of these small BNNTs are strongly dependent on their chirality and diameters. The small zigzag BNNTs become desirable semiconductors and have peculiar distribution of nearly free electron states due to strong hybridization effect. When such a small BNNT is inserted in a larger one, the energy gap of the formed double-walled BNNT can even be much reduced due to the coupled effect of wall buckling difference and interwall NFE-$\pi^*$ hybridization.

---

[a] Corresponding author: wlguo@nuaa.edu.cn

## I. INTRODUCTION

Small-diameter carbon nanotubes (CNTs) exhibit many exotic properties such as anisotropic optical absorption spectra[1] and superconductivity originated from a Peierls distortion.[2,3] These findings have stimulated much interest in study of small nanotubes both theoretically and experimentally.[4-12] Boron nitride nanotubes (BNNTs), predicted by Rubio[13] and then synthesized firstly by Chopra *et al.*[14], have a structural analogy to CNTs. Contrary to the CNTs being metallic or semiconductor depending on chirality, BNNTs are usually believed to be an one-dimensional insulator regardless of their helicity, tube diameters and number of tube walls.[13,15] Of all the properties presented, striking chemical inertness makes the BNNT attractive and prior to the CNT in some environments. Undesirably, the pristine BNNTs are usually believed not suitable for electronic components, though they hold great promise for applications in nanotechnology.[16-18] Herein, we want to stress that such concept possibly stems from little attention on the small-diameter BNNTs. In fact, when the diameter of zigzag BNNT is less than 9.5 Å, the band gap would decrease rapidly with decreasing diameter due to the curvature effect as shown by recent tight-binding[13] and first-principles[19] calculations, which was further confirmed by late GW calculations.[20] This inspires us that when the BNNT enters into small size, a semiconductor with changeable energy gap should be promised to suit for a variety of electronic devices.

In previous study,[19] the (5,0) BNNT with a diameter of 4 Å was claimed as the smallest BNNT stable in free space by total energy comparisons between the BNNT and corresponding flat BN strip. Nevertheless, the total energy comparisons may not be relevant to the question of tube formation as kinetic effects could dominate the growth process. Therefore, the stability of the BNNTs below 4 Å as well as their size dependent properties is worthy of particular investigation. However, compared with the extensive attentions made on small CNTs, detailed understanding of the stability and properties of the small BNNTs is largely lacking. On the other hand, due to the ionicity of B-N bonds, the wall buckling, which is absent in CNTs, is an intrinsic structural feature of BNNTs. Buckling structure can greatly lower the energies of nanostructures[21] or surfaces[22] and consequently enhance their stability. Small BNNTs thereby should possess higher stability than the CNTs with similar diameter as the buckling effect increases rapidly with decreasing tube diameter[23]. From this viewpoint, the study of small BNNTs should be more interesting than those performed with CNT counterparts.

On the other hand, experimental results have shown that small CNTs are usually found inside multi-walled CNTs.[5,7,9] There is, therefore, also a strong motivation to study in detail the stability of small BNNTs inside a larger one, which may shed lights on their experimental synthesis. Besides, the study on double-walled BNNTs (DBNNTs) has demonstrated an interesting variation in their electronic properties when compared with those of freestanding component BNNTs.[24] So it is also important to see the interwall coupling behavior associated with the small BNNTs.

In this article, we present a comprehensive study aimed at predicting the stability and electronic properties of the BNNTs below 4 Å through semi-empirical molecular dynamics (SEMD) simulations and *ab initio* calculations. All the freestanding BNNTs above (2,1) can be stable well over room temperature. Their electronic properties strongly depend on their chirality and corresponding zigzag tubes have become semiconductors, exhibiting distinguished electronic properties from their large insulated family members. The work functions of these small BNNTs also show significant dependence on chirality and diameters. Even the smallest (3,0) BNNT can become globally stable as the core shell of a DBNNT. Surprisingly, the semiconductor energy gap of the small BNNTs can be further remarkably reduced when inserted into an outer BNNT, and the gap reduction sensitively depends on the interwall spacing. These unexpected electronic properties promise the small BNNTs potential applications in electronic and optical devices.

## II. COMPUTATIONAL METHODS AND MODELS

Hartree-Fock-AM1 Hamiltonian[25] is employed in our SEMD simulations to investigate the stability of the small BNNTs, which has been tested to accurately reproduce molecular ground-state geometries, energies and heats of formation. The time step in the SEMD simulations is set to 1 fs. All the simulated BNNTs have length of around 50 Å, and the dangling bonds at both ends of the BNNTs are uniformly terminated by hydrogen atoms. *Ab initio* calculations are carried out to further confirm the stability and evaluate their electronic properties. The ultrasoft pseudo-potentials with a plane-wave basis are employed,[26-28] which is restricted by a cutoff energy of 435 eV. Total energy calculations are performed within both the local density approximation (LDA) and generalized gradient approximation[29] (GGA), while the electronic structure calculations within the LDA only. One-dimensional periodic boundary condition is applied along the nanotube axis. Two adjacent tubes are separated by a vacuum region of at least 14 Å to eliminate the interaction between them. Test calculations show that work functions of the BNNTs are nearly unchanged when the vacuum region increases to 16 Å. The Brillouin-zone integration is sampled by up to 24 special $k$ points for atomic structure relaxation and total 60 $k$ points for electronic structure calculation. The conjugate gradient method is employed to optimize the geometry until the force on each atom is less than 0.01 eV/Å.

## III. STABILITY OF THE SMALL BNNTs

### A. Stability of freestanding BNNTs

Previous theoretical study has shown a critical tube diameter of about 4.0 Å above which the total energies of freestanding BNNTs are lower than those of corresponding planar BN ribbons in free space.[19] Below this size, there are theoretically eight tube models of (2,0), (2,1), (3,0), (3,1), (2,2), (4,0), (4,1) and (3,2). The mechanical stability of these small BNNTs is investigated carefully by performing SEMD simulations with duration up to 3000 fs. It is found that freestanding BNNTs smaller than the (3,0) tube are not stable even at room temperature, which are illustrated by an obvious energy drop during the simulations as shown in Fig. 1(a) for the (2,1) tube. Unexpectedly, the (3,0) BNNT is not only stable at room temperature, but also shows stable total energy throughout the 3 ps SEMD simulation at temperature up to 1600 K,[30] see Fig. 1(a). With further increasing tube diameter, all BNNTs are stable at room temperature regardless of their chirality. For example, the (4,0) BNNT remains mechanically stable up to 2400 K during the SEMD simulations. When compared with CNTs, it is surprising to find by our calculations that the (4,0) BNNT is comparable to a larger (5,0) CNT, which has a diameter of 3.2 Å but can be mechanically stable only to 2500 K. However, the SEMD can only provide a rough estimate for the stability due to the limited tube length and simulation duration.

To confirm the stability of the small BNNTs revealed by the SEMD simulations, we perform careful *ab initio* total energy calculations. In experiment, electron-diffraction analysis has revealed that BN nanotubes favor a zigzag structure in current synthesis processes.[31,32] Therefore, here we concentrate on the (3,0) and (4,0) BNNTs to demonstrate their stability. In order to obtain the approximate energy change path for the BNNT uncurling into a flat strip, the total energy for the optimized structure is calculated as a function of the uncurling angle $\theta$ as defined in the inset of Fig. 1(b). At each uncurling angle, all the atoms are energetically optimized except for the atoms at uncurled edges of the ribbon to determine the system energy. This method is similar to that of Sawada *et al.*[12] and has been successfully used to predict the smallest CNT in free space. The LDA calculations show that when uncurling the (4,0) tube in free space the energy increases with $\theta$, and an energy barrier of 0.22 eV/atom must be overcome to uncurl the tube into a flat ribbon as shown in Fig. 1(b). The energy barrier, which is mainly due to the creation of dangling bonds at the uncurled edges, measures the bond strength of the BNNT. We obtain an equivalent temperature of 1700 K from the energy barrier.[33] The same calculations are also performed with the GGA, which yields an energy barrier of 0.19 eV/atom with an equivalent temperature of 1500 K. For the (3,0) BNNT, relatively lower uncurling energy barriers with equivalent temperatures of 1200 K and 1067 K are obtained by LDA and GGA calculations, respectively.[30] To verify these results, we also perform expensive calculations using the climbing image nudged elastic band method.[34,35] An energy barrier of 0.146 eV/atom is obtained for uncurling the (3,0) into a planar BN ribbon within the LDA, in good agreement with the value present in Fig. 1(b). Generally, the LDA may overestimate the interaction between the B and N atoms while the GGA always makes an underestimation on it.[36] Therefore, the true uncurling energy barrier should fall in range between 1500 K and 1700 K for the (4,0) BNNT as well as between 1067 K and 1200 K for the (3,0) BNNT, some lower than the SEMD predictions. Here, we must be aware from statistical viewpoint that a few atoms may reach the global minimum energy without completely overcoming the energy barriers, leading to lower realistic stability. However, the predicted energy barriers from the SEMD and *ab initio* calculations convince that the freestanding (3,0) BNNT is stable well over room temperature.

### B. Structural properties

The optimized structures of the small BNNTs show a distinct waved structure, with the B and N atoms moving inward and outward, respectively, see the insets in Fig. 2.

We define wall buckling as the difference between the mean radiuses of the cylinders consisting of the N atoms and the B atoms. It is shown that the buckling rapidly increases with decreasing the tube diameter but is nearly independent of the chirality [Fig. 2]. This result is consistent with the reported trend revealed by tight-binding[23] and *ab initio*[37] calculations. The wall buckling is driven by different hybridizations related with the B and N atoms.[15,23] In the acutely curved tube wall, a nonequivalent $sp^3$ hybridization would be mixed on the N atoms due to the existence of isolated pair electrons, which leads to decrease in associated bond angles and form structure similar to that of molecular $NH_3$. The Coulomb repulsion between the isolated pair electrons and bonding electrons makes the N atoms move outward from the tube wall. In contrary, the B atoms move below the wall so that the equivalent planar $sp^2$ hybridization can be maintained to optimize the hybridization energy. Such buckling structure, nearly absent in CNTs, can minimize the curvature energy, and should be the main reason for the high stability of the BNNTs. The strong B-N bond of course is another reason for the distinct stability.

To better understand the effect of tube buckling on the stability of the BNNTs, a comparison between the BNNT and CNT should be helpful. As different chemical compositions are involved in BNNTs and CNTs, the cohesive energy per atom does not provide a suitable measure for the comparison of their relative stability. Here, we define a formation energy $\delta E$ (per atom) for a $B_xN_xC_{1-2x}$ nanotube that is expressed as $\delta E = E_{coh} - x(\mu_B + \mu_N) - (1-2x)\mu_C$, where $E_{coh}$ is the cohesive energy per atom of the $B_xN_xC_{1-2x}$ nanotube, $x$ ($x=0$ for CNTs and $x=0.5$ for BNNTs) is the molar fraction of B and N in the nanotube, and $\mu_i$ ($i$=B, N, C) is the corresponding chemical potential. We choose $\mu_C$ as the cohesive energy per atom of a single graphene sheet and $\mu_B+\mu_N$ as the cohesive energy per BN pair of a hexagonal BN sheet. The definition allows of a direct energy comparison between systems with different chemical compositions.[38-40] Calculated δ$E$ for different nanotubes are listed in Table I. The formation energy of the (4,0) BNNT is close to that of the (5,0) CNT, suggesting a comparable stability between them, consistent with above SEMD simulations. A summary of the results clearly demonstrates higher stability of small BNNTs over the CNTs of similar size and highlights that the buckling structure enhances the stability of the small BNNTs. In fact, additional calculations show that total energy of the (3,0) BNNT is lowered by about 0.13 eV/atom due to the wall buckling. Since the less stable (2,2) CNT has already been synthesized,[7] it is reasonable to expect that, under certain conditions, all the BNNTs larger than (2,1) could be obtained in experiments.

### C. Stability of small BNNTs encapsulated in a larger BNNT

As the total energies of the small BNNTs are higher than those of their uncurled states, examination the behavior of the small BNNTs inside a larger BNNT should be necessary. Firstly, we discuss the energetics with respect to interwall spacing to find the optimal interwall spacing for encapsulating the small BNNTs. Here we concentrate on the DBNNTs (3,0)@($n$,0) and (4,0)@($n$,0) BNNT. The binding energy in the reaction is shown in Fig. 3(a), which is calculated as the energy difference between the total energy of the combined system and the summation of the total energies of two freestanding component nanotubes. The lowest binding energy at $n$ =12 indicates that the (4,0)@(12,0) DBNNT is the optimal zigzag pair, which has an average interwall spacing of 3.12 Å. In the following discussion we will take the (12,0) as the most favorable outer shell for the (4,0) BNNT, but we do not exclude some possible chiral BNNTs as its energetically favorable outer shells. Similarly, our calculations show that the (11,0) BNNT is the most favorable outer tube for the (3,0) BNNT. All the chiral configurations for $n \gtrsim 3$ comply with the optimum rule of $(n,0)@(n+8,0)$ found in normal-sized DBNNTs. Since the experimentally observed interwall spacing in DBNNTs shows large scatter around their optimum value, the (3,0)@(12,0), (4,0)@(13,0), (3,0)@(13,0) and (4,0)@(14,0) DBNNTs are also expected to be obtained in experiments because they are exothermic reactions.

Next, we particularly clarify the stability of the small BNNTs encapsulated in their energetically favorable outer tubes. The binding energy versus the uncurling angle is calculated as shown in Fig. 3(b). It is found that when uncurling the (4,0) tube inside the (12,0) BNNT the binding energy increases monotonically with the uncurling angle, sharply contrasting with that shown in Fig. 1(b). Similar monotonic increase in binding energy during uncurling is also observed in the (3,0)@(11,0) DBNNT. It is apparent that the small BNNTs become globally stable in their energy favorable outer shells. Encapsulation induced enhancement in stability should be able to extend into other small BNNTs. Recent experiments indeed found that freestanding small CNTs are extremely sensitive to the electron beam but can exist stably as the core of a multi-walled CNT.[5,7,9] These results indicate that the small BNNTs are more feasible to be synthesized in a confined space such as in a multi-walled BNNT or in a porous template like zeolite.

## IV. ELECTRONIC PROPERTIES OF THE SMALL BNNTs

### A. Electronic properties of freestanding small BNNTs

The electronic properties of the freestanding small BNNTs are much different from those of their large insulating counterparts. In the pristine (4,0) BNNT, strong $\pi^*$-$\sigma^*$ hybridization results in significant repulsion between the $\pi^*$ and $\sigma^*$ bands. The lowest $\pi^*$ state called singly degenerate state moves close to the Fermi level at the $\Gamma$ point, while the doubly degenerate $\pi$ state is raised at Brillouin zone boundary, forming a indirect energy gap of only 1.8 eV as shown in Fig. 4(a). The gap is greatly reduced from that obtainable in normal-sized BNNTs (~4.3 eV) and approaches the gap width of silicon nanowires.[41] The (3,0) BNNT possesses a smaller direct gap of 1.25 eV. Therefore, the small zigzag BNNT is a standard semiconductor suitable for electronic devices, though a GW correction might widen the energy gap by some degree. Recent theory calculations also reveal a diameter-dependent dielectric function and in zigzag BNNTs, rendering distinct optical properties of the small nanotubes.[42]

In Fig. 4(a), we did not find an eigenstate completely from the nearly free electron (NFE) states, which is a sparse off-tube state existing in normal-sized BNNTs or planar BN sheets.[43] The states have substantially hybridized with the $\pi^*$ state in such an ultrasmall BNNT, making the charge density of $\pi^*$ states not decay even at the tube center [see the inset at bottom of Fig. 4(a)]. The occurred NFE-$\pi^*$ hybridization raises the NFE band to higher energy level, and a large part of the states is meanwhile retreated to the outside of the tube, as shown by the inset at top of Fig. 4(a). As a result, the NFE states have a maximum amplitude at about 2.3 Å out of the (4, 0) tube wall. The out-shifted NFE states of the small BNNT will affect the interwall interaction when it serves as the core shell to form the DBNNT, as will be shown later. These band characters are in sharply contrast with those of the moderate (11,0) BNNT, where the NFE and $\pi^*$ bands are localized at similar energy levels and the NFE states is predominantly confined inside the nanotube as shown in Fig. 4(b).

### B. Chirality dependence of energy gap

In the small BNNTs, a little change in atomic structure may largely affect the overlapping of charge density due to the small tube circumference. Therefore, the electronic structures of the small BNNTs should show sensitive dependence on their chirality. To clarify this important issue, we perform calculations on band structures of the (4,0), (4,1) and (2,2) BNNTs with average diameters of 3.35 Å, 3.7 Å and 2.92 Å, respectively, as shown in Fig. 5(a). It is clear that the electronic structure dramatically changes with chirality. The energy gap increases from 1.8 eV of the (4,0) BNNT to 3.6 eV of the (2,2) BNNT, in sharply contrast with the trend observed in normal-sized BNNTs where the gap is nearly independent of chirality.[13] In spite of the fact that the LDA underestimates the band gap of semiconductors, the chirality dependence of the energy gap shown in Fig. 5(b) should be fairly reliable, as directly supported by recent *ab initio* calculations with self interaction correction.[44]

It can be found that the variation in band gap is mainly due to the shift of the lowest $\pi^*$ state, which roots in the atomic structural feature of the small BNNTs. For convenience, we will call the lowest $\pi^*$ state as singly degenerate state hereafter for all the small BNNTs. In BNNTs, the $\pi^*$ states are mainly distributed on the B atoms from the $p_z$ orbitals perpendicularly to the tube wall. So it can be imagined that a shorter B-B distance along the circumference will be more favorable for an overlapping of the singly degenerate state inside the tube, thus enhancing the $\pi^*$-$\sigma^*$ hybridization. With increasing chiral angle of the BNNT, there are two primary variations that could significantly affect such overlapping: (i) Increasing nearest B-B distance along the circumference. The nearest B-B distance in the (4,0) BNNT is about 2.27 Å, while it increases to 2.317 Å in the (4,1) BNNT and 2.386 Å in the (3,2) BNNT. An increase of the B-B distance of course weakens the overlapping of charge density. (ii) The tilted orientation of the B-B distance with respect to the circumferential direction. This tilting increases with increasing chirality angle as shown in Fig. 6(a), which also disfavors the overlapping inside the tube because the $\pi^*$ states have a parallel alignment along the tube axis. The enhanced overlapping of the singly degenerate state inside the nanotube assists the formation of weak bonds between the B atoms as shown in Fig. 6(b), which determines the downshift of the singly degenerate band.[45] Therefore, zigzag BNNTs always possess the smallest gap within the BNNTs of a given diameter. Besides, we also note that the repulsion between the NFE and $\pi^*$ bands becomes increasingly significant from the (2,2) to (4,0) BNNTs [Fig. 5(a)], indicating that the NFE-$\pi^*$ hybridization becomes stronger with decreasing chiral angle as well. This phenomenon gives us a hint that the amplitude of NFE-$\pi^*$ hybridization is proportional to that of $\pi^*$-$\sigma^*$ hybridization. On the other hand, it is found that the nearest B-B distance along the circumference can also be reduced by decreasing the tube diameter, which is decreased to 2.16 Å in the (3,0) BNNT. That is why the energy gap displays apparent diameter dependence in the small zigzag BNNTs. As expected, since the overlapping between charge density scales as $1/t^2$ where $t$ is the distance between two atoms,[46] the dependences of energy gap on chirality and diameter would become negligible with increasing tube diameter.

### C. Work functions

Nanotube work function is an important factor to quantify the field emission properties from the BNNTs. It has been shown that the BNNTs can show stable, reversible breakdown current at a high bias. However, very few studies have been performed on the work functions of BNNTs,[47] which should exhibit significant change when the BNNTs enter into small size due to the strong orbital hybridization,. Here the work functions of the small BNNTs are calculated and summarized in Fig. 7. The results show that single-walled BNNTs have higher work functions than those of single-walled CNTs, in good agreement with the fact that a higher bias is required to induce the field emission current in BNNTs. The work function change within the calculated diameter range can be up to 0.5 eV, which can be a significant value in device physics. The variation in work functions of the (*n*,*m*) BNNTs versus diameter exhibits three distinct family behaviors (*n*-*m*=3*p*-1, 3*p* and 3*p*+1, *p* is a positive integer), showing strong dependence on chirality. For a given tube diameter, the BNNTs with $n-m=3p+1$ have the highest work functions, while the BNNTs of the other two families enter into a uniform curve. The three families have a common trend that decreases with increasing tube diameter. The family behavior in work function variations is attributed to the different electronic structures stemming from quantized wave vector along the circumference.[48] Due to the large gap value of BNNTs, the gap change induced by this circumferential quantization is not as distinct as that in CNTs, but the quantization is clearly reflected in the change of work functions. On the other hand, the surface dipole and hybridization effects are responsible for the observed diameter dependence of work functions.[49] In contrast with CNTs, the work function of BNNTs hardly increases with increasing tube diameter, mainly because the $\pi^*$-$\sigma^*$ hybridization is more significant than that in CNTs, and overcomes the change from the surface dipole effect.

**D. Electronic properties inside a larger BNNT**

It has been shown that the energy gap of a DBNNT is distinctly reduced from that of the freestanding inner tube.[24] This interesting variation of electronic structure is consistent in the DBNNTs containing the small nanotubes. The calculated energy gap of the (4,0)@(12,0) and (3,0)@(11,0) DBNNTs are calculated to be 1.1 eV and 0.41 eV, respectively, as shown in Figs. 8(a) and (c), significantly reduced from those of the freestanding (3,0) and (4,0) BNNTs. This gap reduction induced by encapsulation of the outer shell is found to be irrespective of the chirality configuration. For the (2,2)@(7,7) DBNNT, the energy gap is reduced by about 0.6 eV from that of a freestanding (2,2) BNNT as shown in Fig. 8(c). The smaller energy gap of the DBNNTs promises them more potential in high-frequency electronic devices and optical applications such as for infrared optical emission. Also, this interesting variation in electronic structures is independent of the number of tube walls as shown in Fig. 8(d), where the (3,0)@(11,0)@(19,0) triple-walled BNNT exhibits similar gap reduction as well. Wave function analysis reveals that the top of valence band is distributed on the outermost BNNT while the conduction band bottom resides on the innermost tube as highlighted in Fig. 8 for all the calculated BNNTs. Apparently, the gap reduction is induced by a misalignment between the energy gaps of the inner and outer BNNTs, whereas the energy gap associated to each component nanotube is nearly unchanged from their freestanding values. Furthermore, we calculate the energy gap as a function of the interwall spacing by varying the tube size from (3,0)@(10,0) to (3,0)@(14,0) as shown in Fig. 9. It is found that the energy gap distinctly decreases with increasing interwall spacing. All the DBNNTs other than the (3,0)@(10,0)[50] have much smaller energy gap than that of the freestanding (3,0) BNNT. The sensitive dependence of energy gap of the DBNNTs on its interwall spacing suggests a new route to efficiently design novel nanodevices. For example, the interwall spacing can be modulated by hydrostatical pressures, and then a tunable gap is available in the DBNNTs.

The mechanism for this interesting variation of electronic structures in DBNNTs is not much clear to date. Here, we show that the interwall NFE-$\pi^*$ hybridization and the buckling difference between the inner and outer walls play crucial roles in determining the energy gap of DBNNTs. Firstly, one should note that there are mainly two kinds of interwall interactions in the DBNNTs within the LDA, namely the $\pi^*$-$\pi^*$ interaction[51] and the NFE-$\pi^*$ hybridization[52], both of which can be dramatically changed by varying the interwall spacing. The $\pi^*$-$\pi^*$ interaction only reacts in the case of small interwall space, such as in the (3,0)@(10,0) DBNNT, which leads to band repulsion between the inner and outer shells. Thus the (3,0)@(10,0) DBNNT has a gap larger than that of the (3,0) BNNT. So we guesses that the gap reduction of the DBNNTs with respect to that of the freestanding inner (3,0) is due to the NFE-$\pi^*$ interaction, which can be understood by the gap change versus interwall spacing [Fig. 9(a)]. In contrast with the $\pi^*$-$\pi^*$ interaction, the NFE-$\pi^*$ interaction can be enhanced by appropriately increasing the interwall spacing. This is because the maximum amplitude of NFE states of a normal BNNT is located at about 1.75 Å inside from the tube wall, while that of the (3,0) tube is at 2.3 Å outside from the tube wall. Thus a interwall spacing of about 4 Å is necessary to hold the maximum amplitude of the NFE states of the both inner and outer tubes, so that the interwall NFE-$\pi^*$ hybridization is maximized for the largest gap reduction induced by encapsulation. In Fig. 9 (a), we indeed observed a critical interwall spacing of about 4 Å, beyond which the energy gap no longer decreases with increasing interwall spacing. Another essential condition for the gap reduction in DBNNTs is the difference between the inner and outer wall buckling. To clarify this issue, we remove the wall buckling of the inner and outer tubes and then recalculate the gap variation versus the interwall spacing. It is surprising to find that the energy gap of the DBNNT is only slightly changed from that of the freestanding inner tube and meanwhile is nearly independent of the interwall spacing [red line in Fig. 9(a)]. As increasing the interwall spacing, the slight gap reduction is just due to the decreased $\pi^*$-$\pi^*$ interaction. Figure 9(b) shows the band structure of the (3,0)@(11,0) DBNNT without of wall buckling, where both the HOMO and LUMO of the DBNNT are from the inner tube. We thus conclude that the peculiar gap reduction in DBNNT roots in the coupled effect between the NFE-$\pi^*$ hybridization and the difference of wall buckling. Since the difference of wall buckling increases with decreasing tube diameter (see Fig. 2), the gap reduction in small DBNNTs is more remarkable than those in normal DBNNTs. For example, the gap of the (3,0)@(11,0) is reduced by 0.5 eV from that of the (3,0) BNNT, while the gap of the (8,0)@(16,0) is only 0.2 eV smaller than that of the (8,0) tube.

The work functions of DBNNTs lie between the work functions of their inner and outer shells. The qualitative feature of the work function variations in DBNNTs can be understood by the charge equilibration model. Generally speaking, when two systems with different work functions come into contact, electron transfer from a lower work function system to a higher work function system takes place in order to equalize the chemical potentials. For the (3,0)@(11,0) DBNNT, the work function of the inner wall is 0.4 eV higher than that of the outer tube. Therefore, a charge transfer from the outer wall to the inner wall should be evident, as confirmed by plotting the corresponding charge redistribution as shown in Fig. 9 (c). The final work function (5.94 eV) of the DBNNT is a linear combination of work functions of the inner (6.15 eV) and outer (5.77 eV) tubes, similar to those in double-walled CNTs.[53,54]

## V. CONCLUSIONS

In summary, we have studied the stability and electronic properties of the small BNNTs with diameter below 4 Å. All the freestanding BNNTs with sizes above (2,1) are guaranteed to be stable well over room temperature. The origin of the high stability of the BNNTs is mainly from the distinct wall buckling minimizing the curvature energy. By inserting these small BNNTs into the interior of a larger BNNT, the small BNNTs can become globally stable. For electronic properties, these small BNNTs can be semiconductors or remain insulators depending on their chirality. Work functions of the small BNNTs are also strongly dependent on their chirality, diameters and number of tube walls, with a variation up to 0.5 eV within the calculated diameter range. Smaller energy gap can be further obtained by inserting them into a larger BNNT as a consequence of the interplay between the NFE-$\pi^*$ hybridization and the wall buckling difference, which also sensitively depends on the interwall spacing of the DBNNT. All these findings on the small BNNTs add different faiths in BN materials for electronic and optical applications.

## ACKNOWLEDGEMENTS

This work was supported by the 973 Program (Contract No. 2007CB936204), the Ministry of Education (Contract Nos. 705021 and IRT0534), National NSF (Contract No. 10732040), and Jiangsu Province Scientific Research Innovation Project for Graduate Student (Contract No. CX07B_064z). We thank Professor C. F. Chen, Professor Z. K. Tang, Professor L. M. Peng, and Professor L. F. Sun for helpful discussions.

[32] M. Terauchi, M. Tanaka, K. Suzuki, A. Ogino, and K. Kimura, Chem. Phys. Lett. **324**, 359 (2000).

[33] The temperature for a stable system can be defined by $\frac{3}{2}N\mathrm{K}_\mathrm{B}T = \sum_{i=1}^{N} E_i$, where $\mathrm{K}_\mathrm{B}$ is the Boltzmann constant, $N$ is the number of atom, $T$ is the temperature and $E_i$ is the kinetic energy of $i^{th}$ atom. Assuming all atoms have the same kinetic energy, when the total kinetic energy $E_t=NE_i$ is higher than the energy barrier $\Delta E$, then we have an so-called critical equivalent temperature to overcome the barrier $3\mathrm{K}_\mathrm{B}T_{eq}/2 = \Delta E$. It should be noted that in a real system, the kinetic energies of atoms have a statistic distribution, and the energy barrier $\Delta E$ can be overcome in certain probability at much lower temperature than the $T_{eq}$. Here the equivalent temperature calculated in such a way only to provide a scale for discussion.

[34] G. Mills, H. Jonsson, and G. K. Schenter, Surf. Sci. **324**, 305 (1995).

[35] G. Henkelman, B. P. Uberuaga, and H. Jonsson, J. Chem. Phys. **113**, 9901 (2000).

[36] A. Marini, P. G. González, and A. Rubio, Phys. Rev. Lett. **96**, 136404 (2006).

[37] L. Wirtz, A. Rubio, R. A. Concha, and A. Loiseau, Phys. Rev. B **68**, 045425 (2003).

[38] T. Dumitricã, M. Hua, and B. I. Yakobson, Phys. Rev. B **70**, 241303(R) (2004).

[39] V. Barone, O. Hod, and G. E. Scuseria, Nano Lett. **6**, 2748 (2006).

[40] O. Hod, V. Barone, J. E. Peralta, and G. E. Scuseria, Nano Lett. **7**, 2295 (2007).

[41] G. Y. Guo and J. C. Lin, Phys. Rev. B **71**, 165402 (2005).

[42] A. J. Read, R. J. Needs, K. J. Nash, L. T. Canham, P. D. Calcott, and A. Qteish, Phys. Rev. Lett. **69**, 1232 (1992).

TABLE I. Calculated total energy $E_{tot}$ and formation energy $\delta E$ of small nanotubes.

| Nanotube | $E_{tot}$ (eV/atom) | $\delta E$ (eV/atom) |
|---|---|---|
| (3,0) CNT | -8.95 | 1.18 |
| (3,0) BNNT | -8.75 | 0.92 |
| (2,2) CNT | -9.04 | 1.09 |
| (4,0) BNNT | -9.14 | 0.53 |
| (5,0) CNT | -9.63 | 0.5 |

FIG. 1. (Color online) Stability of small BNNTs. (a) Total energy per atom of some small BNNTs against released duration by AM1 quantum mechanical based molecular dynamics simulations. A circle is labeled after skipping 10 data points. (b) Total energy per atom of the (3,0) (line with circles) and (4,0) (line with squares) BNNTs obtained by local-density approximation and generalized gradient approximation versus the uncurling angle $\theta$, which is illustrated in the inset where the red (gray) and blue (dark) balls represent the N and B atoms, respectively.

FIG. 2. (Color online) Difference between $d_N/2$ and $d_B/2$ of the equilibrium structures of small BNNTs vs tube diameter. $d_N$ and $d_B$ denote the diameters of the two cylinders formed by the N and B atoms, respectively.

FIG. 3. (Color online) Stability of small BNNTs inside larger BNNTs. (a) Binding energy per atom of the double-walled BNNTs (3,0)@($n$,0) and (4,0)@($n$,0) as a function of the interwall spacing, where $n$ is an integer. (b) Binding energy versus the uncurling angle $\theta$ for the (4,0) tube inside the (12,0) BNNT.

FIG. 4. (Color online) Band structures of the (a) (4,0) and (b) (11,0) BNNTs. The insets show the charge density of nearly free electron (NFE) and the lowest $\pi^*$ states, respectively. The Fermi level is set to zero. The color range maps the value of the charge density, with blue corresponding to the lowest and red to the highest values as illustrated in the right bar.

FIG. 5. (Color online) Dependence of the electronic properties of small BNNTs on chirality. (a) Band structures of the (4,0), (4,1) and (2,2) BNNTs. (b) Energy gap of the small BNNTs as a function of the chiral angle.

FIG. 6. (Color online) Structures (a) and slices of the lowest $\pi^*$ charge density (b) of the small BNNTs with different chirality. The dash line denotes the position along which the charge density is sliced. All the slices cross two nearest B atoms at least.

FIG. 7. (Color online) Work functions of ($n$, $m$) BNNTs. $p$ is an integer.

FIG. 8. (Color online) Band structures of small BNNTs inside larger BNNTs. The band structures of the (a) (3,0)@(11,0), (b) (4,0)@(12,0), (c) (2,2)@(7,7) double-walled BNNTs, (d) (3,0)@(11,0)@(19,0) triple-walled BNNTs. Red lines with hollow squares denote the top valence band and bottom conduction band of the innermost BNNTs while blue lines with hollow circles denote those of the outermost BNNTs.

FIG. 9. (Color online) (a) Energy gap of the (3,0)@($n$,0) with (blue line with squares) and without (red line with bullets) of wall buckling via interwall spacing with $n$ = 10~13. The blue dash and red dot lines denote the band gaps of freestanding (3,0) BNNTs with and without wall buckling, respectively. Axial lattice constants of all the calculated nanotubes are set to 4.25 Å. Thus the (3,0) BNNT is axially stretched, with its energy gap reduced to 0.93 eV from pristine 1.25 eV. (b) Band structure of the (3,0)@(11,0) DBNNT without of wall buckling, and the stipulations follow the convention described in Fig. 8. (c) Charge density plots of charge redistribution in the pristine (3,0)@(11,0) DBNNT [accumulation region: red (solid lines); depletion region: blue (dash lines)].